\documentclass[twocolumn,twoside,slac_two]{revtex4}
\usepackage{graphicx}
\usepackage{fancyhdr}
\begin{document}

\title{Searches for Dark Matter signatures in the Segue~1 dwarf spheroidal galaxy with the MAGIC-I telescope}
\author{S.~Paiano}
\affiliation{Universit\`{a} di Padova and INFN, I-35131 Padova, Italy, and Astronomy Department, Padova, Italy}
\author{S.~Lombardi}
\affiliation{Universit\`{a} di Padova and INFN, I-35131 Padova, Italy}
\author{M.~Doro}
\affiliation{Universitat Aut\`onoma de Barcelona, E-08193 Bellaterra, Spain}
\author{D.~Nieto}
\affiliation{Universidad Complutense de Madrid, E-28040 Madrid, Spain}
\author{\emph{on behalf of the MAGIC Collaboration}}
\affiliation{http://magic.mppmu.mpg.de/}
\author{M.~Fornasa}
\affiliation{Instituto de Astrof\'isica de Andaluc\'ia (IAA-CSIC), E-18008 Granada, Spain}

\begin{abstract}
Despite the interest in Dark Matter (DM) searches is currently more focused 
on underground experiments, a signature of DM annihilation/decay in gamma-rays 
from the space would constitute a smoking gun for its identification. Here, 
we present the results of the survey of Segue~1 by the MAGIC-I telescope 
performed in 2008 and 2009. This source is considered by many as the most DM 
dominated Milky Way satellite galaxy known so far. The nearly 43 hours of data 
taken constitute the deepest observation ever made on a single dwarf galaxy by 
Cherenkov telescopes. No significant gamma-ray emission was found above an energy 
threshold of $100$~GeV. Integral upper limits on the gamma-ray flux were calculated 
assuming various power-law spectra for the possible emission spectrum and for 
different energy thresholds. We also discuss a novel analysis that fully takes into 
account the spectral features of the gamma-ray spectrum of specific DM models in 
a SuperSymmetric scenario.
\end{abstract}

\maketitle

\thispagestyle{fancy}

\section{Introduction}
In the $\Lambda$CDM cosmological scenario about 80$\%$ of the matter of 
the Universe is believed to be composed of non-baryonic matter, called 
Dark Matter (DM). The most popular DM candidates are the WIMPs (weakly 
interacting massive particles) supposed to be cold, electrically neutral, 
stable, and massive~\cite{spergel}. Among the huge plethora of WIMP 
candidates, the best motivated ones are related to the SuperSymmetrical 
(SUSY) and Extra Dimensional extensions of the Standard Model of particle 
physics~\cite{bertone}.\\
In the Minimal SuperSymmetric extension of the Standard Model (MSSM), the 
neutralino $\chi$ represents an excellent cold DM candidate with a relic 
density compatible with the WMAP bounds. Since the neutralino is a Majorana
particle, pairs of $\chi$ can annihilate into Standard Model particles, e.g., 
quarks, leptons, and W bosons. The subsequent hadronization of those particles
results in a continuum emission of gamma-rays characterized by a cut-off at 
the neutralino mass and by possible spectral features like bumps or a hardening 
of the spectral slope.\\ 
The expected gamma-ray flux from DM-annihilating astrophysical objects, as 
function of the energy threshold $E_0$ and the integration region $\Delta\Omega$, 
within which the signal is integrated, can be factorized in two terms:
\begin{equation}\label{eq:flux}
\Phi(>E_0,\Delta\Omega)=\Phi^{PP}(>E_0)J(\Delta\Omega).
\end{equation}
The so-called particle physics factor $\Phi^{PP}$ depends on the features 
of the DM particle, and can be written as:
\begin{equation}\label{eq:part}
\Phi^{PP}(>E_0)=\frac{1}{4\pi}\frac{<\sigma_{ann}v>}{2 m^{2}_{\chi}}\int^{m_{\chi}}_{E_{0}}\sum^{n}_{i=1}B^i \frac{dN^{i}_{\gamma}}{dE}dE,
\end{equation}
where $<\sigma_{ann}v>$ is the velocity averaged annihilation cross-section, 
and $B^i$ is the particular branching ratio for the $i$-th annihilation channel.\\
The term J$(\Delta\Omega)$ (the so-called astrophysical factor) is given by 
the line--of--sight integral over the DM density squared within a solid angle
$\Delta\Omega$, and depends on the density profile of the DM halo of the source:
\begin{equation}\label{eq:astro}
J(\Delta\Omega)=\int_{\Delta\Omega}\int_{los}\rho^2(r(s,\Omega))dsd\Omega.
\end{equation}\\
Since the gamma-ray flux of DM annihilation is proportional to the square of 
the DM density, only sources with high expected DM densities are good targets 
for DM indirect searches. Among these, the dwarf spheroidal satellite galaxies
(dSphs) of the Milky Way (MW) are interesting objects thanks to their relative 
proximity to the Earth, to their high mass--to--light ratio (with values within 
tens and thousands of M$_{\odot}$/L$_{\odot}$) and to 
the expected absence of conventional gamma-ray sources within the system~\cite{sanchez,gilmore}. 
So far, around two dozen dSphs have been identified. Segue~1, discovered in 2006 by the SDSS~\cite{segue}, 
is located at 28~kpc from the Galactic Center, at (RA,DEC)=(10.12$^h$, 16.08$^{\circ}$). 
Kinematics studies applied to 66 member stars allowed to estimate its mass--to--light 
ratio to be in the range 1320-3400 M$_{\odot}$/L$_{\odot}$~\cite{segueML}, highlighting Segue~1 
as the most DM dominated dSph known so far.\\
The MAGIC-I telescope is a 17~m dish Imaging Atmospheric Cherenkov Telescope (IACT), 
located at the Roque de los Muchachos Observatory, in the Canary Island of La Palma 
(2200 m a.s.l.). Thanks to its low energy threshold ($\sim$60~GeV at Zenith), high 
flux sensitivity ($1.8\%$ of the Crab Nebula flux in 50~hour of observations above 
$\sim$250~GeV), and good angular and energy resolution ($0.1^{\circ}$ and $30\%$ respectively, 
at $100$~GeV)~\cite{crab}, MAGIC-I is a suited instrument for the indirect search for DM candidates with energy of the 
neutralino mass or the Kaluza-Klein state. 
%
\section{MAGIC-I observation and analysis results}
\begin{figure}[h]
\centering
\includegraphics[width=70mm]{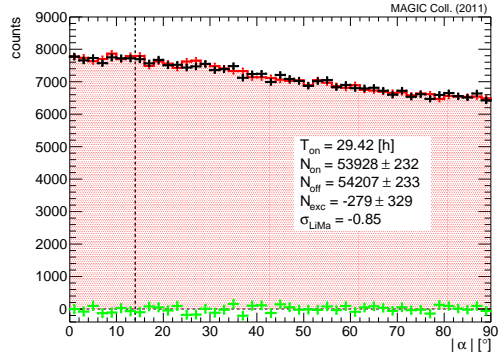}
\caption{$\alpha$-plot from 29.4 hours of Segue~1 observation above $100$~GeV. 
Red points represent the signal (ON distribution), black points the background 
(OFF distribution), and green points their difference. The vertical dashed line 
at $\alpha=14^{\circ}$ is the fiducial region below which the excess event number 
is estimated. Image taken from ~\cite{segue1paper}.}
\label{alphaplot}
\end{figure}

\begin{figure}[h]
\centering
\includegraphics[width=70mm]{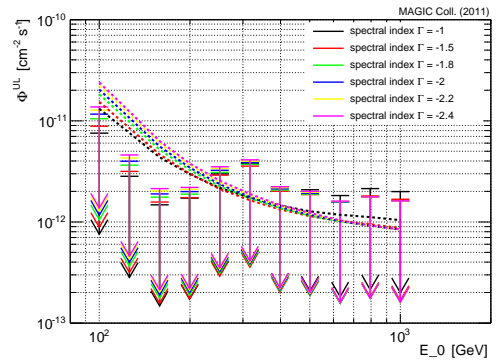}
\caption{Integral flux ULs from Segue~1. The arrows indicate the integral flux upper 
limits for different power-law spectra and energy thresholds. The dashed lines indicate 
the corresponding integral ULs if zero significance $\sigma_{Li,Ma}$ is assumed. Image taken from ~\cite{segue1paper}.}
\label{intUL}
\end{figure}

A search for a possible DM gamma-ray signal coming from Segue~1 was performed by 
the MAGIC-I telescope between November 2008 and March 2009, for a total of 29.4~hours 
of observation time (after data selection). The data analysis was performed using 
the standard MAGIC-I analysis and reconstruction software~\cite{bretz}. The number 
of gamma-ray candidates events from the direction of the source was estimated using 
the distribution of $|\alpha|$ angles, which are related to the orientation of the showers. 
The overall analysis cuts were optimized and cross-checked for point-like sources with the 
aid of contemporaneous Crab Nebula data. 
In Figure~\ref{alphaplot}, the $|\alpha|$-plot above $100$~GeV is shown. The number of excess
events was computed in a fixed fiducial signal region with $|\alpha|<14^{\circ}$ and 
resulted to be $N_{exc}(> 100~\mbox{GeV})=-279\pm329$, corresponding to a significance 
of $-0.85\sigma$, computed using eq.(17) of Li$\&$Ma~\cite{LiMa}.
Since results were consistent with no signal over the background, we derived Upper Limits 
(ULs) on the flux, calculated using the Rolke method~\cite{rolke} at $95\%$ confidence level, 
and assuming a $30\%$ systematic uncertainty. Figure~\ref{intUL} shows the integral ULs achieved 
by the MAGIC-I observation of Segue~1 considering different energy thresholds $E_{0}$ and 
different power-law spectra with spectral index $\Gamma = -1,-1.5,-1.8,-2,-2.2,-2.4$.
It is worth noting that, using the Rolke method, the ULs on the number of the excess events, 
and consequently the integral flux ULs, are affected by statistical fluctuations quantified
by the significance of the observation $\sigma_{Li,Ma}$. 
This is an intrinsic feature of the statistical method exploited in the analysis 
and it should be taken into account when comparing ULs from different analyses.
To show this effect, in Figure~\ref{intUL} we plot also the ULs (dashed lines) computed 
assuming a value for $\sigma_{Li,Ma}$ equal to zero (with number of ON events 
equal to the number of OFF events in the signal region of $|\alpha|$-plot) for different 
values of spectral index and energy threshold.

\section{Constraints on Dark Matter models}
\begin{figure}[]
\centering
\includegraphics[width=70mm]{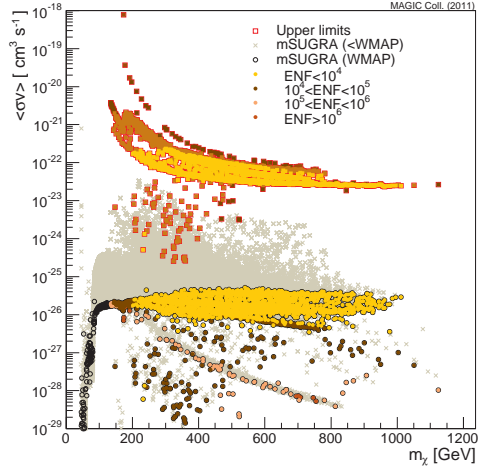}
\caption{Velocity averaged annihilation cross-section ULs from Segue 1 MAGIC-I data 
computed for individual points in the scan. Grey crosses indicate 
the velocity averaged annihilation cross-section value for those points in the scan 
that pass the SM constraints and with a relic density lower than 
WMAP bound. The full circles only consider models within $3\sigma_{WMAP}$ 
from WMAP bounds. For each of these full circles the UL on the cross-section can be computed from the Segue 1 data (after energy threshold 
optimization) and it is indicated here by a square.
Circles and squared are colored in term of the enhancement factor. Image taken from~\cite{segue1paper}.}
\label{DM_ULs}
\end{figure}

Assuming a particular form for Segue~1 DM halo, and a given particle model 
for the DM candidates, we can translate the integral ULs derived from the 
Segue~1 observation into constraints on the DM annihilation rate.\\ 
Motivated by results from cosmological simulation, the DM halo around Segue~1 
was modeled by using the Einasto radial profile~\cite{essig}
with $\sigma_s=$1.1$\times$10$^8$~M$_{\odot}$~kpc$^{-3}$, r$_s$=0.15~kpc, and 
n=3.3. With those parameters, the total astrophysical factor of 
Segue~1 results to be $J(\Delta\Omega)=1.78\times$10$^{19}$~GeV$^2$~cm$^{-5}$~sr. 
Since the analysis was performed assuming point-like source cuts (corresponding to 
an angular integration of 0.14$^\circ$ above $100$~GeV), we estimated the effective 
astrophysical factor within the analysis cuts to be used in the following analysis,
being its value $\widetilde{J}(\Delta\Omega)=1.14\times$10$^{19}$~GeV$^2$~cm$^{-5}$~sr 
(corresponding to the 64$\%$ of the total astrophysical factor).\\
Concerning the particle physics, we restricted ourselves to the case of a SUSY model 
in which the presence of a discrete symmetry (R--parity) guarantees that the Lightest 
SuperSymmetric Particle (LSP) is stable over cosmological timescales and, therefore, 
a good DM candidate. We considered a 5-dimensional subspace of the MSSM called mSUGRA~\cite{cham}, 
for which the basic parameters are the universal masses of the gauginos ($m_{1/2}$) 
and scalars ($m_0$), the trilinear coupling ($A_0$), the ratio of the vacuum expectation 
values of the two Higgs fields (tan$\beta$) and the sign of the Higgsino mass term (sign$(\mu)$). 
In order to study the phenomenology of mSUGRA we performed a grid scan over the parameter space, 
for a total of 5$\times$10$^6$ points (for the details see \cite{segue1paper}). \\
The full circles of Figure~\ref{DM_ULs} represent all the models of the scan \textit{i)} where 
the lightest SUSY particle is a neutralino, \textit{ii)} that survive the Standard Model 
constraints and \textit{iii)} with a relic density compatible with the value derived by WMAP 
data within three times its experimental error $\sigma_{WMAP}$~\cite{wmap}. 
For each DM model of the scan, we computed the integral flux UL (above an energy 
threshold E$_0$), using the Segue~1 data and the specific gamma-ray spectrum derived from the individual 
DM model. Since the spectra for each DM model have different shapes and cut-offs, the value of the optimal
energy threshold E$_0$ was computed individually for each DM mass. 
We then converted the flux ULs into ULs on the velocity averaged cross-section to have a direct comparison 
of experimental data with the theoretical predictions. The results are plotted in Figure~\ref{DM_ULs}
as function of the neutralino mass: each DM model of the scan (full circle) is compared to its own UL (square).
For each point we defined an enhancement factor (ENF) as the ratio between the UL on the velocity averaged cross-section 
and the value predicted by mSUGRA. This quantity quantifies how far away we are from excluding some portions of 
the mSUGRA parameters space. From Figure~\ref{DM_ULs} it can be seen that ENFs for model compatible with the WMAP 
bounds are typically above 10$^{3}$, while typical values are of the order of $10^{4-5}$.
\section{Impact on PAMELA preferred region}
\begin{figure}[]
\centering
\includegraphics[width=70mm]{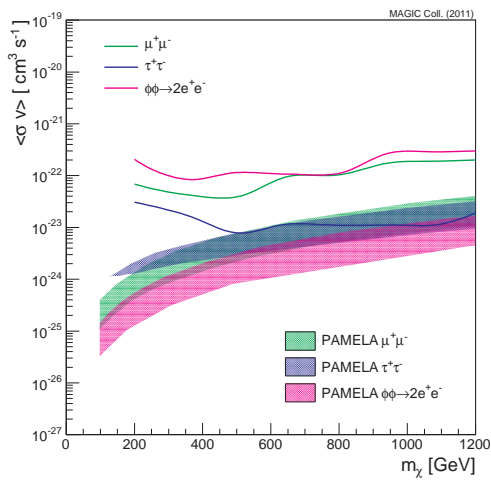}
\caption{Exclusion lines for a neutralino DM annihilating exclusively 
into $\mu^+\mu^-$ (green lines) or $\tau^+\tau^-$ (blue line), and for 
a DM candidate interacting with a light intermediate state $\phi$ decaying
into a pair of electrons (pink line). The same annihilation channels 
(with the same color coding) are considered to draw the regions in the plane 
that provide a good fit to the PAMELA measurement of the energy spectrum 
of the positron fraction. Image taken from ~\cite{segue1paper}.}
\label{Pamela_ULs}
\end{figure}

We tested our ULs on some of the models proposed in the literature that can explain the PAMELA data~\cite{pamela} 
for the energy spectrum of the positron fraction $e^+/(e^++e^-)$ as due to DM annihilation into leptons.
The regions in the $(m_{\chi},<\sigma_{ann} v>)$ plane that provide a good fit to the PAMELA measurements~\cite{pamelamisura, cholis2} 
for three different channels of DM annihilation are shown in Figure ~\ref{Pamela_ULs}. The annihilation channels 
$\chi\chi \to \mu^+\mu^-$, $\chi\chi \to \tau^+\tau^-$ have been taken from SuperSymmetry, while for 
$\chi\chi \to \phi^+\phi^- \to 2e^+e^- $ the existence of a new dark force, mediated by the carrier $\phi$ that 
decays into leptons~\cite{arhani78}, has been assumed. In Figure ~\ref{Pamela_ULs} we plot the ULs obtained from the 
Segue~1 data, using again the specific DM annihilation spectra. We can see that, in this case, the ENFs needed to 
meet the PAMELA-favoured region are much smaller than those found for mSUGRA, and in the case of annihilation 
into $\tau^+\tau^-$ our ULs are probing the relevant regions. However, it is worth mentioning that, since the 
uncertainty in the Segue~1 astrophysical factor is quite large \cite{essig}, an improvement in accuracy of astrophysical factor 
value could be able to put more stringent constrains and to confirm the exclusion of the PAMELA region for DM particle 
annihilating in $\tau^+\tau^-$.
\section{Conclusions}
A search for a possible DM gamma-ray signal coming from Segue~1 was performed by the MAGIC-I telescope. 
No hints of signal were found above the background for energies larger than $100$~GeV. Integral ULs 
on the gamma-ray emission were computed assuming different power-law energy spectra. Within the mSUGRA 
scenario, a large scan of neutralino models was performed over the parameter space. Subsequently for 
each simulated DM model, the ULs on the velocity averaged annihilation cross-section $(m_{\chi},<\sigma_{ann}v>)$  were derived 
separately for each point in the scan in order to account for the dependence on the specific spectra. 
Results indicate that a general exclusion plot cannot be drawn to constrain the parameter space, so we 
provide the results in terms of enhancement factors. A minimum boost on the flux is found of the order of 
10$^3$ (for models compatible with WMAP) while the typical values are at 10$^{4-5}$. 
MAGIC-I data of Segue~1 can be useful to put constraints on those DM models that are provided in the literature 
to explain the PAMELA data. Our ULs are probing the PAMELA region for the DM models annihilating into $\tau^+\tau^-$ 
but the robustness of this result could be improved, decreasing the uncertainty in the astrophysical factor. \\
Although the MAGIC-I observation did not result in a detection, and the ULs require still high 
flux enhancement factors to actually match the experiment sensitivity, an analysis like the one presented here 
is able to point out details and features that can be important for future deep exposures of this or similar objects, 
with next-generation Cherenkov experiments.
%
\bigskip 
\begin{acknowledgments}
\footnotesize{
We would like to thank the Instituto de Astrof\'{\i}sica de
Canarias for the excellent working conditions at the
Observatorio del Roque de los Muchachos in La Palma.
The support of the German BMBF and MPG, the Italian INFN, 
the Swiss National Fund SNF, and the Spanish MICINN is 
gratefully acknowledged. This work was also supported by 
the Marie Curie program, by the CPAN CSD2007-00042 and MultiDark
CSD2009-00064 projects of the Spanish Consolider-Ingenio 2010
programme, by grant DO02-353 of the Bulgarian NSF, by grant 127740 of 
the Academy of Finland, by the YIP of the Helmholtz Gemeinschaft, 
by the DFG Cluster of Excellence ``Origin and Structure of the 
Universe'', by the DFG Collaborative Research Centers SFB823/C4 and SFB876/C3,
and by the Polish MNiSzW grant 745/N-HESS-MAGIC/2010/0.}
\end{acknowledgments}



\begin{thebibliography}{9}   


\bibitem{spergel}
D.~N.~Spergel \textit{et al.}, ApJS \textbf{170} (2007) 377.
\bibitem{bertone}
G.~Bertone, D.~Hooper, J.~Silk, Phys. Rept. \textbf{405} (2005) 279-390.
\bibitem{sanchez}
M.~Sanchez-Conde \textit{et al.}, AIP Conf. Proc.\textbf{1166} (2009) 191-196.
\bibitem{gilmore}
G.~Gilmore  \textit{et al.}, Astrophys.J.\textbf{663} (2007) 948-959.
\bibitem{segue}
V.~Belokurov \textit{et al.} [SDSS Colaboration], Astrophys.J. \textbf{654} (2007) 897-906 [astro-ph/0608448].
\bibitem{segueML}
J.D.~Simon \textit{et al.}, (submitted) (2010) astro-ph.GA/1007.4198.
\bibitem{crab}
J.~Albert \textit{et al.} [MAGIC Collaboration], Astrophys.J. \textbf{674} (2008) 1037-1055 [arXiv:0705:3244].
\bibitem{bretz}
T.~Bretz \textit{et al.}, Proc. 29th ICRC 2005, Tsukuba.
\bibitem{LiMa}
T.P.~Li, Y.Q.~Ma, Astrophys.J. \textbf{272} (1983) 317-324.
\bibitem{segue1paper}
J.~Aleksi\'c \textit{et al.} [MAGIC Collaboration], JCAP \textbf{06} (2011) 035.
\bibitem{rolke}
W.~A.~Rolke, A.~M.~Lopez, J.~Conrad, Nucl.Instrum.Meth. \textbf{A551} (2005) 493-503 [physics/0403059].
\bibitem{essig}
R.~Essig, N. Sehgal, L. E. Stringari, M. Geha, J.D. Simon, Phys.Rev \textbf{D82} (2010) 123503 [astro-ph/1007.4199].
\bibitem{cham}
A.~H.~Chamseddine, R.~L.~Arnowitt, P.~Nath, Phys.Rev.Lett \textbf{49} (1982) 970.
\bibitem{wmap}
E.~Komatsu \textit{et al.} [WMAP Collaboration], Astrophys.J.Suppl. \textbf{192} (2011) 18 [astro-ph.CO:1001.4538].
\bibitem{pamela}
O.~Adriani \textit{et al.} [PAMELA Collaboration], Nature \textbf{458} (2009) 607-609 [astro-ph/0810-4995].
\bibitem{pamelamisura}
I.~Cholis \textit{et al.}, JCAP \textbf{0912} (2009) 007 [astro-ph/0810.5344].
\bibitem{cholis2}
I.~Cholis \textit{et al.},  Phys.Rev \textbf{D80} (2009) 123518 [astro-ph/0811.3641].
\bibitem{arhani78}
N.~Arkani-Hamed \textit{et al.} Phys.Rev \textbf{D79} (2009) 015014 [hep-ph/0810.0713].













\end{thebibliography}

\end{document}